\documentclass[12pt, draftclsnofoot, journal, letterpaper, onecolumn]{IEEEtran}

\usepackage{cite}
\usepackage{amsmath,amsthm,amssymb,amsfonts}
\usepackage{algorithm}
\usepackage{algorithmic}
\usepackage{graphicx}
\usepackage{color}

\usepackage[normalsize]{subfigure}

\newcommand{\mv}[1]{\mbox{\boldmath{$ #1 $}}}

\begin{document}

\title{Exploiting Interference for Secrecy Wireless Information and Power Transfer}
\author{Yuan Liu, Jie Xu,  and Rui Zhang
\thanks{Y. Liu is with the School of Electronic and Information Engineering, South China University of Technology, Guangzhou 510641, China (e-mail: eeyliu@scut.edu.cn).}
\thanks{J. Xu is with the School of Information Engineering, Guangdong University of Technology, Guangzhou 510006, China (e-mail: jiexu@gdut.edu.cn).}
\thanks{R. Zhang is with the Department of Electrical and Computer Engineering, National University of Singapore, Singapore 117583 (e-mail: elezhang@nus.edu.sg).}
}

\maketitle

\vspace{-1.5cm}

\begin{abstract}
Radio-frequency (RF) signals enabled wireless information and power transfer (WIPT) is a cost-effective technique to achieve two-way communications and at the same time provide energy supplies for low-power wireless devices. However, the information transmission in WIPT is vulnerable to the eavesdropping by the energy receivers (ERs). To achieve  secrecy communications with information nodes (INs) while satisfying the energy transfer requirement of ERs, an efficient solution is to exploit a dual use of the energy signals also as useful interference or artificial noise (AN) to interfere with the ERs, thus preventing against their potential information eavesdropping. Towards this end, this article provides an overview on the joint design of energy and information signals to achieve energy-efficient and secure WIPT under various practical setups, including simultaneous wireless information and power transfer (SWIPT), wireless powered cooperative relaying and jamming, and wireless powered communication networks (WPCN). We also present some research directions that are worth pursuing in the future.
\end{abstract}

\section{Introduction}
Future wireless networks are expected to constitute billions of low-power wireless devices (such as sensor nodes, radio frequency (RF) identification (RFID) tags and Internet-of-things (IoT) devices) for diversified  applications, and it is crucial to provide them with satisfactory communication quality of service (QoS), guaranteed data security, and sustainable energy supply. RF signals enabled wireless information and power transfer (WIPT) has been recently recognized as a promising technique to achieve two-way communications  and provide cost-effective energy supplies for low-power wireless devices at the same time \cite{KrikidisCM,BiCM}. In general, there are mainly two types of applications for WIPT, namely simultaneous wireless information and power transfer (SWIPT) and wireless powered communication networks (WPCN), as shown in Figs. 1 and 2, respectively. In SWIPT, hybrid access point (H-AP) simultaneously broadcasts information and energy signals to communicate with information nodes (INs) and power energy receivers (ERs) in the downlink; while in WPCN, H-AP broadcasts energy signals to power both ERs and INs in the downlink, and INs use the harvested energy to transmit information back to H-AP in the uplink \cite{BiCM}.

Despite the technology advancements, WIPT systems face new data security challenges, since the information transmission of the INs (in both downlink and uplink) is vulnerable to be intercepted by the ERs that are untrusted and can be potential eavesdroppers \cite{ChenNg}. In SWIPT systems, ERs are normally located much closer to the H-AP than INs due to their different requirement of receiver power sensitivity (e.g., $-$10dBm for energy harvesting versus $-$60dBm for information reception) \cite{LiuTSP}. Due to this ``near-far'' effect, untrusted ERs can easily overhear and eavesdrop the downlink information intended to INs. In WPCN, both ERs and INs are located close to the H-AP to harvest the RF energy in the downlink. Therefore, INs' transmitted information in the uplink is easy to be eavesdropped by nearby untrusted ERs.

\begin{figure}[t]
\begin{centering}
\includegraphics[scale=1.5]{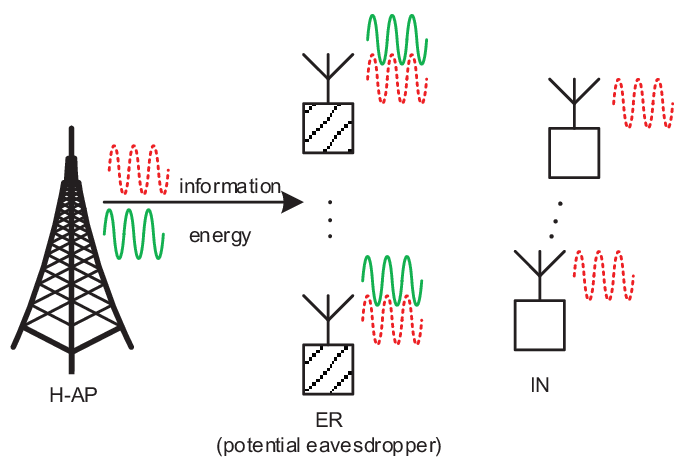}
\vspace{-0.1cm}
 \caption{An example of SWIPT, where the information sent to INs in the downlink is vulnerable to be eavesdropped by ERs.}\label{fig:SISO}
\end{centering}
\vspace{-0.3cm}
\end{figure}
\begin{figure}[t]
\begin{centering}
\includegraphics[scale=1.2]{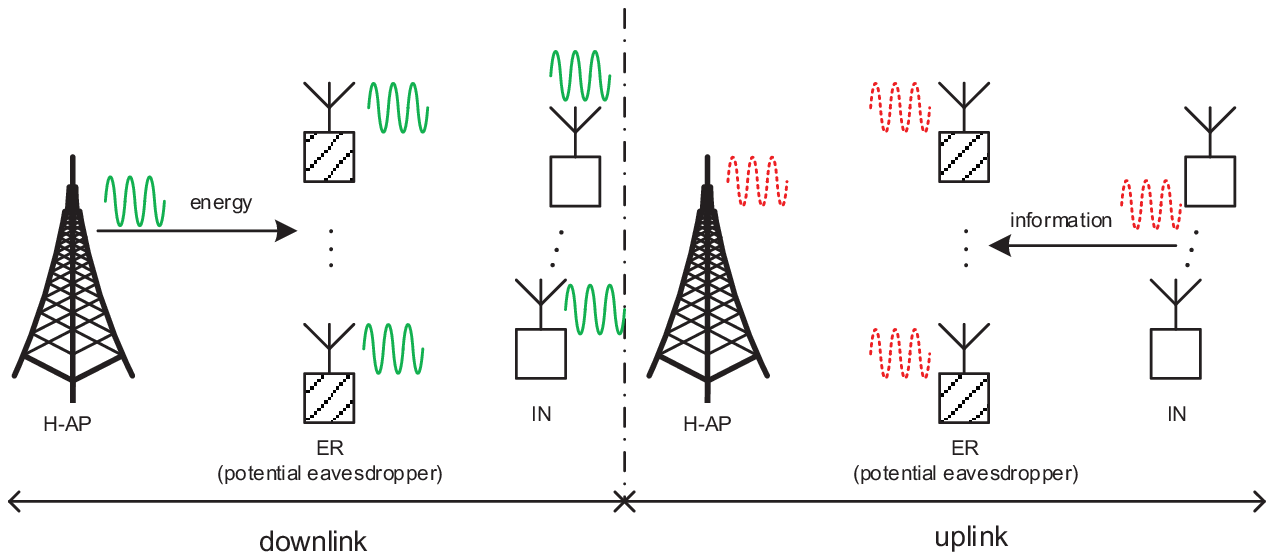}
\vspace{-0.1cm}
 \caption{An example of WPCN, where the information transmission of INs in the uplink is easy to be overheard by nearby ERs.}\label{fig:wpc}
\end{centering}
\vspace{-0.3cm}
\end{figure}

Physical-layer security has been recognized as a promising technique to secure wireless communications against malicious eavesdropping attacks. 
The objective of physical-layer security is to maximize the secrecy rate of a communication channel, which corresponds to the achievable data rate of this channel provided that the eavesdropper cannot intercept or decode any information. In the literature, there are various approaches that have been proposed to improve the secrecy rate. Among them, the artificial noise (AN) (see, e.g., \cite{GoelNegi2008}) and cooperative jamming (see, e.g., \cite{DongHan2010}) based designs are appealing, where properly designed AN or jamming signals are transmitted by the transmitter itself (together with the confidential message) and by the external helping nodes (HNs), respectively, for the purpose of interfering with the malicious eavesdroppers to improve the secrecy rate.

In this article, we integrate the physical-layer security in  WIPT  to overcome the challenging information leakage problem due to  ER eavesdroppers. Such secure WIPT systems aim to achieve two-way secrecy communications with INs while satisfying the energy harvesting requirements at ERs. We present an efficient solution by exploiting {a dual use of the energy signals in WIPT, for not only energy transfer, but also as useful interference or AN to jam ERs against their potential information eavesdropping.} With such consideration, we provide an overview on the joint design of information and energy signals under various practical system  setups, by investigating two types of IN receivers (namely, Type-I and Type-II IN receivers) that can or cannot cancel the energy signals (or AN) before decoding the information, respectively. {It is assumed that the H-AP accurately knows the channel state information (CSI) to the receivers.} The outline of this article is given as follows.

\begin{itemize}
  \item First, we consider AN-aided secrecy SWIPT systems. We show that when there is only a single antenna at the H-AP, due to the ``near-far'' effect, secrecy communications are only feasible for Type-I IN receivers with AN cancelation employed. On the other hand, when there are multiple antennas at the H-AP, a joint beamforming design of information and energy signals can help achieving secrecy communications while ensuring the energy harvesting requirements, for both Type-I and Type-II IN receivers.
  \item Next, we consider the secrecy SWIPT with wireless powered cooperative relaying and jamming to improve secrecy communication performance. In this approach, trusted idle ERs in the system are enabled as external HNs, which can use the harvested energy from the H-AP to help relay information to the intended IN receivers and jam the untrusted ER eavesdroppers in the {\it downlink}, with a joint design of both information and energy signals.

  \item Furthermore, we also address  the secrecy WPCN with downlink energy transfer and uplink secrecy communications. In this case, wireless powered cooperative jamming is used to exploit trusted ERs as HNs to help for interfering with the {\it uplink} eavesdropping of suspicious ERs. {It is crucial to efficiently schedule these trusted ERs with joint downlink energy transfer and uplink jamming design, in order to optimize the uplink secrecy communication performance. However, this problem has not been addressed before, to the authors' best knowledge.}
\end{itemize}
Along with the above discussions, we also point out some promising directions for future work in both secrecy SWIPT and secrecy WPCN. Finally, we conclude this article.

\section{Secrecy Communication in SWIPT: An Artificial Noise Approach}

In this section, we consider the downlink secrecy SWIPT as shown in Fig. \ref{fig:SISO}, where a single H-AP serves multiple INs and ERs at the same time, { by sending them either information or energy}. In the following, we focus on a special setup with one single-antenna IN and one single-antenna ER to draw essential insight, in which  two cases with single-antenna and multi-antenna H-AP are considered.

\subsection{The Case with Single-Antenna H-AP}

%
%
%

To start with, we consider the simplest case with one single-antenna H-AP serving one single-antenna IN and one single-antenna ER. For the purpose of illustration, let $h_I$ and $h_E$ denote the channel power gains from the H-AP to the IN and the ER, respectively, where $h_I < h_E$ holds to be consistent with their ``near-far" locations. Also, let $s_0$ (with unit power) denote the confidential information signal to be sent to the IN, and $P$ denote the transmit power of the H-AP. Conventionally, the H-AP transmits with only information signal (which also conveys RF energy), by setting the transmit signal as $\sqrt{P}s_0$. In this case, the amount of power harvested at the ER is $\eta h_E P$, where $0<\eta\leq1$ denotes the energy conversion efficiency at the ER. As for the secrecy communication to IN,  the received signal power at the (far) IN (i.e., $h_I P$) is always weaker than that at the (near) ER (i.e., $h_E P$). In this case, any information decodable at the IN can always be decoded or eavesdropped by the ER. Therefore, transmitting solely information signals is  infeasible to achieve secrecy communication in the single-antenna SWIPT setup.

To overcome this security problem caused by the ``near-far'' issue, it is crucial to employ an AN approach by additionally sending a dedicated energy signal or AN to jam the ER eavesdropper; furthermore, we need to enable the IN receiver to pre-cancel the AN before decoding the information  (as will be shown in detail later), in order to achieve non-zero secrecy rate.
%
%
%
In the AN approach, the H-AP splits the transmit power $P$ into two components, with a fraction of $(1-\alpha)$  to send  the information signal $s_0$ for the IN and the remaining  fraction of $\alpha$  to send  the dedicated energy signal or AN (denoted by $s_1$ with unit power) for the ER, where $0 \le \alpha \le 1$ denotes the transmit power splitting ratio.  In this case, the transmitted baseband signal at the H-AP is expressed as $x=\sqrt{P(1-\alpha)}s_0+\sqrt{P\alpha}s_1$. Accordingly, the amount of power harvested at the ER is $\eta P(1-\alpha)h_E+\eta P\alpha h_E= \eta Ph_E$, which is regardless of  the transmit power splitting ratio $\alpha$. As for the secrecy communication,  we consider two types of IN receivers (namely  Type-I and Type-II IN receivers), which can and cannot cancel the AN before decoding the information signal, respectively. In order to enable Type-I IN receiver to perform AN cancellation, it needs to know the AN used at the H-AP  {\it a priori}, based on a practical  physical-layer key distribution method  described as follows \cite{HongTVT16}. First, both the H-AP and the IN should pre-store a sufficiently large ensemble of sequences that are used as the seeds to generate Gaussian distributed AN, and the index of each sequence in the ensemble is regarded  as a ``key''. Then, the H-AP randomly picks up one sequence and transmits its index confidentially to the IN before data transmission. Accordingly, the H-AP is able to generate a random AN signal that is only known to the IN but unknown at the ER. This is due to the fact that without the knowledge of the employed key and given a very large key set, the ER cannot decode the AN even  with a long-term observation, since  the complexity is practically too large and thus infeasible. With the AN known at both the H-AP and the IN, the AN cancellation is thus implementable at the Type-I IN receiver.

Under the Type-I and Type-II IN receivers, the received signal-to-interference-plus-noise ratios (SINRs) are expressed as $\gamma_{I}^{(\rm I)} = \frac{P(1-\alpha)h_I}{\sigma^2}$ and $\gamma_{I}^{(\rm II)} = \frac{P(1-\alpha)h_I}{P\alpha h_I + \sigma^2}$, respectively, and the SINR at the ER eavesdropper is given as $\gamma
_{E} = \frac{P(1-\alpha)h_E}{P\alpha h_E+\sigma^2}$, where $\sigma^2$ denotes the noise power at the receivers of both the IN and the ER. By assuming that  $s_0$ and $s_1$ are independent Gaussian  random variables, the achievable secrecy rate at the Type-$i$ IN receiver is expressed as \cite{GoelNegi2008}
\begin{equation}\label{eqn:1}
R_s^{{(i)}}=\left[\log_2\left(1+\gamma_{I}^{(i)}\right) -
\log_2\left(1+\gamma_{ E}\right)\right]^+,
\end{equation}
where $i\in\{{\rm I},{\rm II}\}$ and $[x]^+=\max\{x,0\}$. {The achievable secrecy rate $R_s^{{(i)}}$ is a non-concave function with respect to the transmit power splitting ratio $\alpha$. To maximize $R_s^{{(i)}}$, we can adopt a one-dimensional search over $0\le \alpha \le 1$.}

\begin{figure}[t]
\begin{centering}
\includegraphics[scale=1]{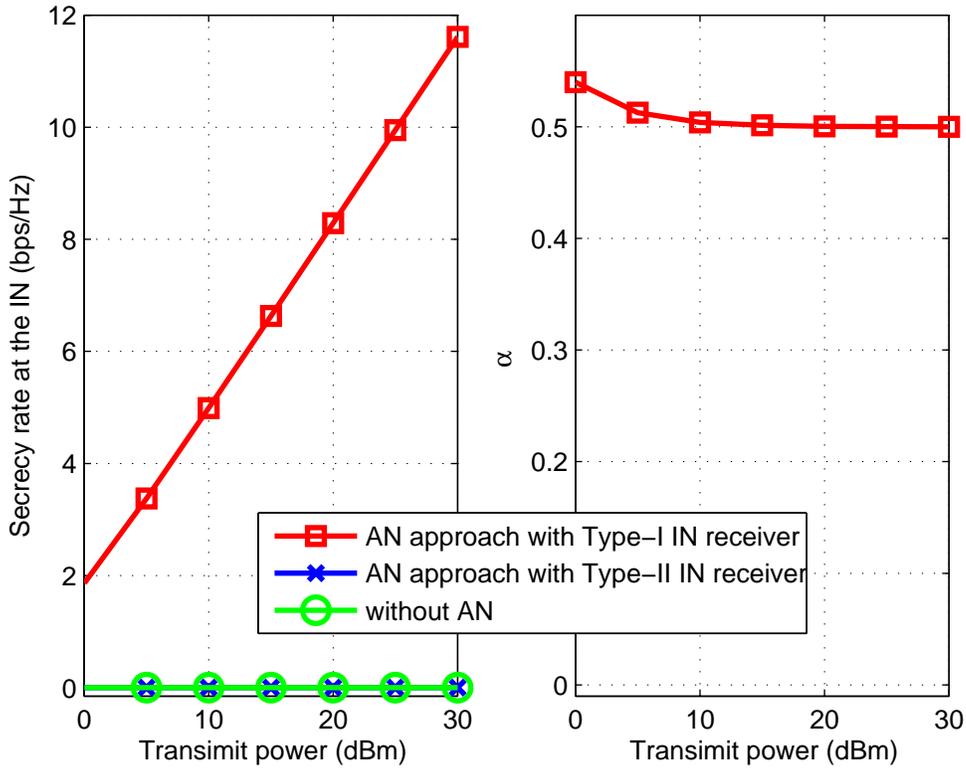}
\vspace{-0.1cm}
 \caption{Secrecy rates versus the transmit power at the H-AP in a SWIPT system with a single-antenna H-AP. }\label{fig:siso-p}
\end{centering}
\vspace{-0.3cm}
\end{figure}

To illustrate the necessity of the AN approach and the AN cancellation at the (Type-I) IN, {we conduct simulations  to compare the performance of the two AN approaches with Type-I and Type-II IN receivers, versus the conventional approach without AN. The results are shown in Fig. \ref{fig:siso-p}. In the conventional approach, the achievable secrecy rate corresponds to $R_s^{{(\rm I)}}$ or $R_s^{{(\rm II)}}$ in \eqref{eqn:1} with $\alpha = 0$.} In this simulation, the distances from the H-AP to the IN and the ER are $d_I=20$ meters (m)  and $d_E=2$m, respectively, the path loss exponent is assumed to be $3$, and the noise power is $-80$dBm. {Under this setup with ``near-far'' deployment, we have $h_I < h_E$.} Note that in all schemes, the harvested powers at the ER are identical, {and thus are not shown. The left subfigure of Fig. \ref{fig:siso-p} shows the maximum secrecy rate versus the transmit power $P$. It is observed that the secrecy rates of the conventional approach without AN and the AN approach with Type-II IN receiver are always zero. This is due to the fact that as $h_I < h_E$, we have $\gamma_{I}^{({\rm II})} < \gamma_{E}$ irrespective of the transmit power splitting ratio $\alpha$. As a result, for both schemes, the received SINR at the IN receiver is always smaller than that at the ER eavesdropper, and hence, the secrecy rate is always zero. By contrast, it is observed that the secrecy rate achieved by the AN approach with Type-I IN receiver is strictly positive and monotonically increasing with the transmit power. This result is expected, and shows that under the singe-antenna H-AP case, the AN approach is only beneficial in improving the secrecy rate when AN cancellation (i.e., Type-I IN receiver) is employed, due to  the ``near-far" deployment. The right subfigure of Fig. \ref{fig:siso-p} shows the optimal transmit power splitting ratio $\alpha$ in the AN approach with Type-II IN receiver versus the transmit power $P$.} It is observed that the optimal $\alpha$ converges to about 0.5 as transmit power increases, i.e., the transmit power is equally split to the information signal and the energy/AN signal. {This is because when the transmit power $P$ goes to infinity, we can approximate the secrecy rate as $R_s^{({\rm I})}\rightarrow\log_2\left(\frac{P(1-\alpha)h_I}{\sigma^2}\right)-\log_2\left(1+\frac{1-\alpha}{\alpha}\right)=\log_2\left(\frac{P(1-\alpha)\alpha h_I}{\sigma^2}\right)$, for which the maximum is attained when $\alpha = 0.5$.}


%

In the literature, the  benefit of using AN at the H-AP and employing AN cancelation at the IN has been further exploited under other system setups. For example, the authors in \cite{HongTVT16} considered  fading channels, where the H-AP adaptively adjusts the power assigned for information and AN signals over different fading states to optimally exploit the channel dynamics for improving the average secrecy rate. Furthermore, the authors in \cite{YuanTWC16} investigated a more general orthogonal frequency-division multiple access (OFDMA) scenario with multiple INs and ERs, where all other receivers (i.e., ERs and other INs) are potential eavesdroppers for each IN. In \cite{YuanTWC16},  the H-AP adds independent AN over each sub-carrier and only the intended IN at that sub-carrier can {\it a-priori} know  the corresponding key to  cancel the AN before decoding its information.

\subsection{The Case with Multi-Antenna H-AP}

Multi-antenna beamforming has  been recognized as an efficient technique to improve both communication rates and energy transfer efficiency in SWIPT \cite{ZhangHo2013}, by steering RF signals towards targeted INs/ERs with focused information and/or energy beams. Similarly, we can exploit the benefit of multi-antenna H-AP in secrecy SWIPT systems. {Suppose that there are $N>1$ antennas at the H-AP, and denote the $N \times 1$ channel vectors from the H-AP to the IN and the ER as $\mv h_I$ and $\mv h_E$, respectively. Let $\mv w_{I}$ and $\mv w_{E}$ denote the $N\times 1$ unit-norm transmit beamforming vectors at the H-AP for the information signal and the energy signal (or AN), respectively. Under the Type-I and Type-II IN receivers, the received SINRs are re-expressed as $\gamma_{I}^{(\rm I)} = \frac{P(1-\alpha)|\mv h_I^H \mv w_{I}|^2}{\sigma^2}$ and $\gamma_{I}^{(\rm II)} = \frac{P(1-\alpha)|\mv h_I^H \mv w_{I}|^2}{P\alpha |\mv h_I^H \mv w_{E}|^2 + \sigma^2}$, respectively, and the SINR at the ER eavesdropper is given as $\gamma_{E} = \frac{P(1-\alpha)|\mv h_E^H \mv w_{I}|^2}{P\alpha |\mv h_E^H \mv w_{E}|^2+\sigma^2}$, where the superscript $H$ denotes the conjugate transpose of a vector. Then, the achievable secrecy rate at the Type-$i$ IN receiver, $i\in\{\rm{I},\rm{II}\}$, is given by $R_s^{{(i)}}$ in \eqref{eqn:1} by substituting the newly defined SINRs.}

Consider first the conventional design without AN, {for which the achievable secrecy rate corresponds to $R_s^{{(\rm I)}}$ or $R_s^{{(\rm II)}}$ with $\alpha = 0$}. In this case, a positive secrecy rate is achievable if the H-AP transmits the information beam {$\mv w_I$} lying in the null space of the channel vector {$\mv h_E$} to the ER. This is in sharp contrast to the case with single-antenna H-AP, where the secrecy rate is always zero without AN employed. As a result, we expect that by additionally exploiting the gain provided by the AN approach, both the secrecy rate and energy transfer efficiency can be further improved via proper information and energy/AN beamforming design.


In general, the design of information and energy/AN beams critically depends on the types of IN receiver, and obtaining the optimal beamformers corresponds to solving complicated optimization problems. Instead of considering the optimal solution, we adopt the following intuitive designs to draw insights. The H-AP first sets the  information beam {$\mv w_I$} following the maximum-ratio transmission (MRT) principle based on the channel vector {$\mv h_I$} to the IN, and then designs the energy/AN beamforming vector {$\mv w_{E}$} depending on the type of IN receiver considered.  When Type-I IN receiver is used with the AN cancellation, the H-AP designs the energy/AN beam {$\mv w_{E}$} following the MRT principle based on the channel vector  {$\mv h_{E}$} to the ER, so as to maximally transfer energy and also maximally degrade the ER's received SINR. When Type-II IN receiver is considered without AN cancellation, the energy beam {$\mv w_{E}$} should be designed based on a zero-forcing (ZF) principle to lie in the null space of the IN's channel vector {$\mv h_{E}$}, so as to avoid the interference to the IN receiver. By exploiting the spatial degrees of freedom brought by the multiple antennas at the H-AP, positive secrecy rates are achievable for both types of IN receiver, which is different from the case with a single-antenna H-AP, where positive secrecy rate is achievable only for Type-I IN receiver.

\begin{figure}[t]
\begin{centering}
\includegraphics[scale=1]{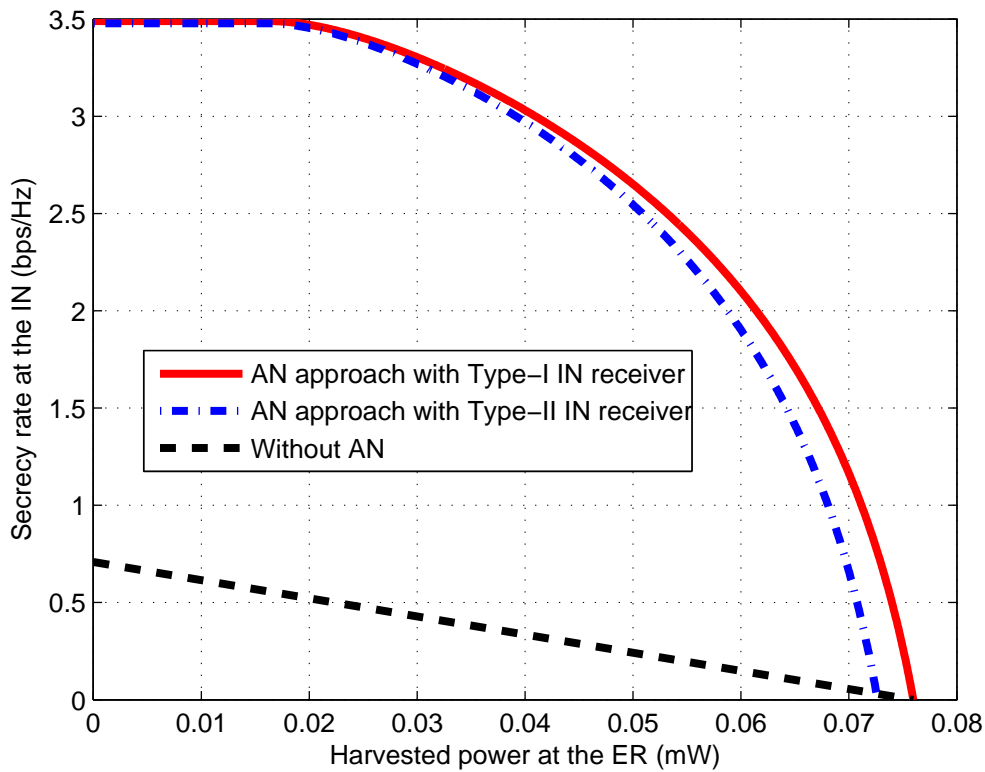}
\vspace{-0.1cm}
 \caption{The secrecy rate at the IN versus the harvested power at the ER for the secrecy SWIPT system in the case with multi-antenna HAP.}\label{fig:mimo-region}
\end{centering}
\vspace{-0.3cm}
\end{figure}

To compare the performances of the AN approach under Type-I and Type-II IN receivers, Fig. \ref{fig:mimo-region} provides a numerical example to show the secrecy rate at the IN versus the harvested power at the ER, where the system parameters are set same as those in the case with single-antenna H-AP, except that the number of transmit antennas at the H-AP is $N = 4$ with the directions from the H-AP to the IN and the ER are specifically set as 0 and 60 degrees, respectively.
For comparison, we also consider the conventional design without AN, in which the H-AP can either transmit over the ER channel's null space to achieve secrecy communication to the IN but without any energy delivered to the ER, or use the MRT beamforming to the ER for maximizing its harvested energy but without any confidential information conveyed. We use time-sharing between the two strategies to achieve both positive secrecy rate and positive harvested energy in general.
From Fig. \ref{fig:mimo-region}, it is observed that both the conventional design without AN and the AN approach with Type-II receiver  achieve positive secrecy rates, which is different from the case with single-antenna H-AP in Fig. \ref{fig:siso-p}, where the secrecy rate is always zero. This is obtained by properly designing the information and energy/AN beamforming to exploit  the additional  spatial degrees of freedom. Furthermore, it is observed that the AN approach with Type-I IN receiver achieves  the best performance in terms of both secrecy rate and  harvested power, thanks to the exploitation of both the cancelation of AN and the beamforming gain.

%

In the literature, the authors in \cite{LiuTSP} and \cite{NgTWC} considered multi-antenna  secrecy SWIPT systems with one ``far" IN as well as multiple ``near" ER eavesdroppers, by considering Type-II and Type-I IN receivers, respectively, where information and energy/AN beamforming vectors are jointly determined following the similar design principle above. {For future work, how to extend the design for secrecy SWIPT to the case with multiple INs each with one or more antennas is still an open problem, which, however, is challenging to solve. On one hand, with multiple INs, the H-AP in general needs to design multiple transmit information beams to different INs, in order to properly control the inter-user interference among them, in addition to ensuring the secrecy performance. On the other hand, the use of multi-antennas at both the H-AP and INs leads to matrix optimization problems that are more difficult to solve than the above example problem with vector variables only.}

\section{Secrecy SWIPT with Wireless Powered Cooperative Relaying and Jamming}
%
%

\begin{figure}[t]
\begin{centering}
\includegraphics[scale=1.8]{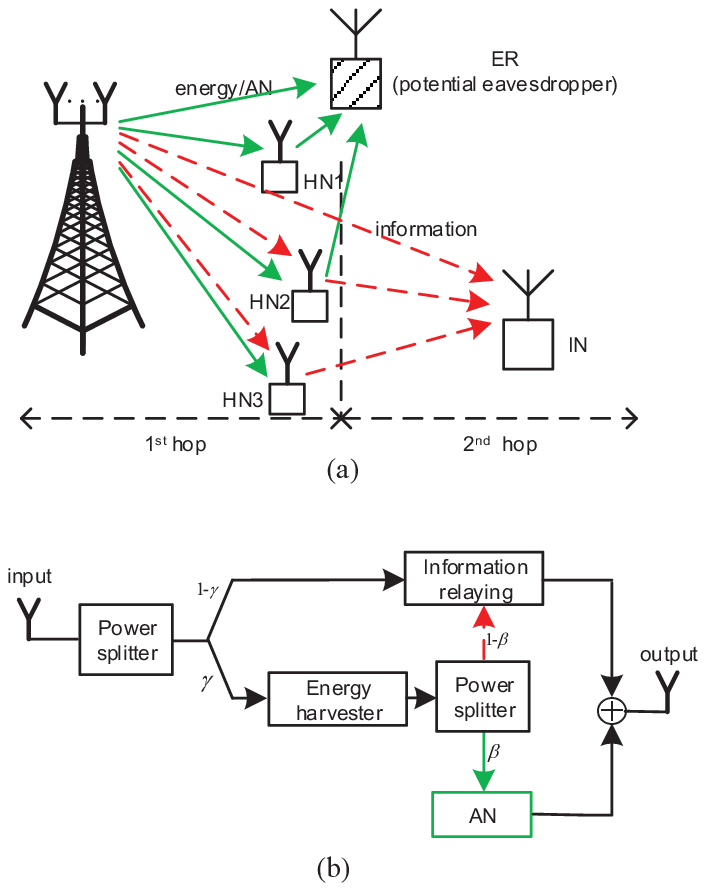}
\vspace{-0.1cm}
 \caption{Secrecy SWIPT system with wireless-powered cooperative relaying and jamming: (a) System model; (b) Operation at  each wireless-powered helping node (HN).}\label{fig:relay}
\end{centering}
\vspace{-0.3cm}
\end{figure}

This section considers another appealing solution named wireless powered cooperative relaying and jamming to improve the secrecy rate of IN in secrecy SWIPT systems. In future, practical wireless networks will constitute numerous low-power wireless devices, and as a result, it is practical that during the IN's communication, some trusted and idle ERs are located between the H-AP and INs. This motivates the idea of wireless powered cooperative relaying and jamming to improve the performance of secrecy SWIPT, where these trusted ERs are enabled  as friendly HNs in both relaying the information from the H-AP to the IN, and sending AN to cooperatively jam the untrusted ER eavesdroppers. As these HNs are of low power, to avoid their energy waste, their individual energy consumption for relaying and jamming should be no larger than that harvested from the H-AP.

%
%
%
%
%

In particular, an example secrecy SWIPT system with wireless powered cooperative relaying and jamming is shown in Fig. \ref{fig:relay}(a), which can be implemented based on the  ``harvest-then-relay-and-jam'' protocol consisting of two time slots: in the first slot, the H-AP sends confidential information signals to both INs and HNs, and transmits energy/AN signals to charge HNs and ERs as well as interfere with ER eavesdroppers at the same time; in the second slot, HNs use the harvested energy to relay information to INs, where AN signals are also sent to defend against the ERs' eavesdropping.

In order to harvest energy as well as relay information and jam with AN, the HNs should adopt new transceiver architectures, an example of which is shown in Fig. \ref{fig:relay}(b), operating as follows. In the first time slot, the HN uses a receive power splitter to  split its received RF signal into two parts, one with a fraction of power $\gamma$ for harvesting the energy to be used in the second time slot and the other with the remaining portion of power $1-\gamma$ for receiving the information to be relayed. In the second slot, the harvested energy at the HN is further split into two parts by a ratio $0\leq\beta\leq1$, with a portion of $\beta$ used for generating AN to jam ER eavesdroppers and the other portion $1-\beta$ for relaying the received information.  Specifically, in a particular time instant, each HN can adjust its operation among the following modes by adjusting the receive and transmit power splitting ratios $\gamma$ and $\beta$.
\begin{itemize}
\item \emph{Harvest-then-jam:} The HN uses all its harvested energy to jam ER eavesdroppers by setting transmit and receive power splitting ratios as  $\beta=1$ and $\gamma = 1$, respectively.
\item \emph{Harvest-then-relay:} The HN uses all its harvested energy to relay confidential information by setting $\beta=0$ and $0<\gamma<1$.
\item \emph{Harvest-then-relay-and-jam:} The HN uses its harvested energy for dual purposes of relaying and jamming by setting $0<\gamma<1$ and $0<\beta<1$.
\end{itemize}
For example, intuitively, when the HN is near ER eavesdroppers but far from the INs (e.g., HN1 in Fig. \ref{fig:relay}(a)), it may choose to operate in the harvest-then-jam mode; when the HN is near INs but far from ER eavesdroppers (e.g., HN2 in Fig. \ref{fig:relay}(a)), it may work in the harvest-then-relay mode; when the HN is near both (e.g., HN3 in Fig. \ref{fig:relay}(a)), it can work in the general harvest-then-relay-and-jam mode.

Furthermore, when the HN needs to relay confidential information (i.e., with $0<\beta<1$ and $0\le\gamma<1$), it can choose its relaying protocol between amplify-and-forward (AF) and decode-and-forward (DF), depending on whether the key (for AN design) is
known a-priori at the HN or not (i.e., Type-I or Type-II receiver of HN). Intuitively, when the key is known {\it a-priori}, at the receiver side HNs can cancel the AN from the H-AP for more efficient DF confidential message relaying; furthermore, at the transmitter side, different HNs can coordinate their respective AN signal design to achieve beamforming (so-called coordinated beamforming), such that the AN will be coherently combined at the ER eavesdroppers  for more efficient jamming. In this case, a significantly improved secrecy communication performance would be achieved, which is at the cost of more implementation complexity, due to the requirement of key sharing (for AN cancellation) from the H-AP to HNs, as well as the decoding processing for extracting the AN at each individual HN. Instead, when the HNs do not know the key (for AN design) {\it a-priori}, it is generally desirable to use AF relaying to avoid the decoding operation at HNs. This also prevents HNs from overhearing the confidential information. 


%

To optimize the performance of the secrecy SWIPT with wireless powered cooperative relaying and jamming, it is crucial to perform a network-wide optimization to determine the operation mode and relaying protocol of different HNs, jointly with their transmit and receive power splitting ratios, as well as the transmit power allocation/beamforming for information and energy/AN signals at both the H-AP and HNs. Such an optimization is also dependent on the types of receivers at the HNs/INs and subject to the energy harvesting constraints at HNs (i.e., the relaying and jamming energy consumption cannot exceed the harvested energy).
{Due to the above issues, the performance optimization for the secrecy SWIPT with wireless powered cooperative relaying and jamming is a difficult problem to solve, and even developing a general problem formulation is a challenging task.} Instead of addressing the general scenario, in the literature there have been several works \cite{Xing2015,Xing2016,LiuZhou2016,BiChen2016} considering various specific setups. For example, the authors in \cite{Xing2015,Xing2016} considered the case with wireless powered AF relaying, where the ER eavesdroppers can only overhear the confidential information relaying from HNs in the second time slot. The optimal AF relaying processing and AN beamforming at HNs are jointly designed to maximize the sum secrecy rate. In \cite{LiuZhou2016,BiChen2016}, the authors considered the case with only wireless powered jamming, where the H-AP first powers the HN and then the HN jams the eavesdropper by using the harvested power.

{Despite the above progress, the extension to more general setups still requires substantial future work. For example, in practice ER eavesdroppers that are not far away from both the H-AP and HN may be able to combine the signals overhead from both hops to degrade the system's secrecy performance, thus making the secrecy design more challenging. Moreover, the multiple wireless-powered HNs may operate in different modes as aforementioned, i.e., harvest, relay, and/or jam. How to select their optimal modes under general setups are also  challenging  problems unsolved.}

\section{Secrecy Communication in WPCN}

Besides secrecy SWIPT, another important application of WIPT is secrecy WPCN as shown in Fig. \ref{fig:wpc}, where INs use the harvested energy from the H-AP to send confidential information back to the H-AP, in the presence of untrusted ER eavesdroppers. To the best of our knowledge, how to improve the secrecy communication performance for WPCN has not been addressed in the literature yet.

Similar as the conventional WPCN without security consideration in  \cite{JuTWC}, the secrecy WPCN can be generally implemented in a ``harvest-then-transmit'' protocol, where the transmission is divided into two time slots: one for wireless energy transfer from the H-AP to a set of INs as well as ERs, and the other for confidential messages transmission from the INs back to the H-AP. In this case, the uplink information can be easily eavesdropped because untrusted ERs (potential eavesdroppers) may be located near the H-AP for energy harvesting.
%
To improve the secrecy rate from INs to the H-AP via defending against untrusted ERs' eavesdropping, an efficient solution is to employ wireless powered cooperative jamming similarly as in the previous section, in which trusted idle ERs in the network are employed as friendly HNs to jam untrusted ERs' uplink eavesdropping (instead of downlink eavesdropping in the previous section).

Specifically, as HNs are not aware of the INs' transmitted confidential information messages, they will operate in the harvest-then-jam mode to interfere with the untrusted ER eavesdroppers, with $\beta=1$ and $\gamma = 1$ in Fig. \ref{fig:relay}(b). In order for more efficient jamming, different HNs should share the same key for AN design, such that they can use coordinated beamforming to maximize the jamming power to ER eavesdroppers. This also simplifies the AN cancellation at the H-AP, as only one key is used. Furthermore, the maximization of uplink secrecy rates requires a joint scheduling for the downlink wireless energy transfer of the H-AP, the uplink information transmission of the INs, and the uplink jamming of the HNs, subject to the energy harvesting constraints at both the INs and HNs. For instance, allocating a longer time slot for wireless energy transfer can lead to higher transmit power of INs and higher jamming power of HNs, but this in turn reduces the time for confidential message transmissions from HNs to the H-AP. An efficient time allocation between the two slots is thus crucial.




The other issue faced in the secrecy WPCN system is the so-called ``doubly near-far'' problem, where ``far'' INs (i.e., INs far away from the H-AP) would harvest less wireless energy in the downlink but needs more transmit power in the uplink to achieve the same communication rate as ``near'' INs. Furthermore, due to the limited available power at far INs, far INs are more vulnerable to be eavesdropped by ERs than near INs. To tackle the ``doubly near-far" problem in secrecy WPCN, it is efficient to enable nearby HNs to help relaying the confidential information to the H-AP. In order to implement such a wireless powered cooperative relaying and jamming in secrecy WPCN, one additional time slot for information relaying is required. In this case, the HN can be implemented based on the structure in Fig. \ref{fig:relay}(b) as follows: in the first time slot, the HN sets the receiver power splitting ratio to be $\gamma = 1$ for harvesting the wireless energy; in the second time slot, the HN sets $\gamma = 0$ for receiving the confidential information from far INs; and in the third time slot, the HN splits its transmit power into two parts for information relaying and sending AN, respectively. A more sophisticated time allocation among the three slots together with the joint downlink and uplink scheduling is necessary to fully reap the gain of wireless powered cooperative relaying and jamming in this case.

{
In addition, backscatter WPCN has recently emerged as a new type of WPCN by leveraging the technique of backscatter communications, where a wireless device without active RF components can reflect back (rather than actively radiate) the RF signals received from the H-AP for the purpose of communications. Here, the reflected signals are modulated via the device by properly controlling the mismatch between the antenna and load impedance. While most existing works focused on using the reflected signals for delivering information, how to use such signals as AN for jamming to improve the secrecy performance, or even use them for dual goals of AN and information signal at the same time is an interesting topic that has not been addressed yet.
}

\section{Concluding Remarks}

This article provided a new  perspective in improving the secrecy communication performance in emerging WIPT systems while ensuring the energy transfer requirements, which exploits a dual use of energy signals as useful interference or AN to combat against potential eavesdropping by untrusted ERs. In particular, we discussed three secrecy WIPT setups, namely, SWIPT, wireless powered cooperative relaying and jamming, and WPCN, respectively. For each setup, we presented the joint information and energy/AN signals design by considering two types of IN receivers that can or cannot cancel the energy/AN signals. We also discussed the design challenges and some future research directions. {Furthermore, there are other interesting issues that are unaddressed in this article due to space limitation, which are briefly discussed in the following.

The implementation of the secrecy WIPT requires the H-AP to accurately know the channel state information (CSI) to both INs and ERs, which is practically a difficult task, especially for the CSI to untrusted ER eavesdroppers. For the purpose of initial investigation, this article assumes that the ER eavesdroppers are existing energy users in the network, and thus they are willing to cooperate in helping the H-AP obtain their CSI, for the purpose of facilitating the energy transfer. However, if the ER eavesdroppers intend to eavesdrop information rather than receiving energy, then it is difficult for the H-AP to acquire their actual CSI. How to design the secrecy WIPT in this case is thus more challenging.

Furthermore, the proposed AN approach with Type-I IN receiver relies on the assumption that the keys for generating the AN can be shared secretly between the secrecy transmitters and receivers. In practice, the secrecy keys may be overheard by ER eavesdroppers. In this case, how to achieve secrecy WIPT with only partially secure key exchange is another challenging open problem.}


\bibliographystyle{IEEEtran}
\bibliography{IEEEabrv,Secrecy}

\end{document}